\documentclass[12pt,tightenlines,eqsecnum,floats,aps,amsmath,amssymb,tightenlines,nofootinbib,prd,shownopacs,notitlepage]{revtex4}

\usepackage{graphicx,verbatim}
\usepackage{amsmath}
\usepackage{amsfonts}
\usepackage{amssymb}
\usepackage[colorinlistoftodos]{todonotes}
\usepackage{epstopdf}

\newcommand{\pb}[1]{\hbox{\lower0.5ex\hbox{${}_{\leftarrow}$}}\kern-1.9ex{#1}}
\def\chi{\theta}

\def\be{\begin{equation}}
\def\ee{\end{equation}}
\def\ba{\begin{eqnarray}}
\def\ea{\end{eqnarray}}

\def\h{\hat}

\def\f{\frac}

\def\M{\mathbb{M}}

\def\rmd{\mathrm{d}}
\def\S{\mathcal{S}}

\def\H{\mathcal{H}}

\def\hphi{\hat{\phi}}
\def\hPhi{\hat{\Phi}}
\def\stress{\langle \hat{T}_{ab}(x) \rangle_{\rm ren}}
\def\phisq{\langle \hat{\phi}^2(x)\rangle_{\rm ren}}
\def\biphi{\langle \h\phi(x)\,\h\phi(\xp)\rangle}

\def\go{\mathring{{g}}}
\def\phio{{\phi}^\circ}
\def\biphio{\langle \h\phio(x)\,\h\phio(\xp)\rangle}

\def\hphio{{\hat{\phi}^\circ}}

\def\hPhi{\h\Phi}

\def\vx{\vec{x}}

\def\vxp{\vec{x}^{\,\prime}}
\def\etap{\eta^{\,\prime}}

\def\xp{x^{\,\prime}}

\begin{document}

\title{Space-like Singularities of General Relativity:\\ A Phantom menace?}

\author{Abhay Ashtekar${}^{1}\,$}
\email{ashtekar.gravity@gmail.com} \author{Adri\'an del R\'io${}^{1}\,$}
\email{axd570@psu.edu}
\author{Marc Schneider ${}^{1,2,3}\,$}
\email{mschneid@sissa.it}
\affiliation{${}^{1}$ Institute for Gravitation and the Cosmos \& Physics Department,\\ The Pennsylvania State University, University Park, PA 16802 U.S.A}
\affiliation{${}^{2}$ Scuola Internazionale Superiore di Studi Avanzati (SISSA)\\ 
via Bonomea, 265, Trieste 34136 ITALY}
\affiliation{${}^{3}$ INFN Sezione di Trieste, via Valerio 2, 34127 Trieste, Italy}

\begin{abstract}

The big bang and the Schwarzschild singularities are space-like. They are generally regarded as the `final frontiers' at which space-time ends and general relativity breaks down. We review the status of such space-like singularities from three increasingly more general perspectives. They are provided by (i) A reformulation of classical general relativity motivated by the Belinskii, Khalatnikov, Lifshitz conjecture on the behavior of the gravitational field near space-like singularities; (ii) The use of test quantum fields to probe the nature of these singularities; and, (iii) An analysis of the fate of these singularities in loop quantum gravity due to quantum geometry effects. At all three levels singularities turn out to be less menacing than one might a priori expect from classical general relativity. Our goal is to present an overview of the emerging conceptual picture and suggest lines for further work. In line with the \emph{Introduction to Current Research} theme, we have made an attempt to make it easily accessible to all researchers in gravitational physics.  

\end{abstract}

\maketitle

\section{Introduction}
\label{s1}

One of the first papers that Professor Padmanabhan --or Paddy, as he was known in the community-- wrote as a Ph.D. student was \emph{``Quantum conformal fluctuations in a singular space-time"}, co-authored with his advisor Jayant Narlikar. \cite{tpjvn}. First two sentences of the abstract set the stage for the subsequent discussion: ``The cosmological solutions of Einstein's general relativistic equations lead inevitably to space-time singularities. However, general relativity is only an approximation to a fully quantized theory of gravity and we need to consider whether singularity persists in the quantum domain." The paper goes on to argue that quantum considerations imply that the classical prediction cannot be trusted. In honor of Paddy's memory, we will provide a contemporary perspective on this issue, drawing on results that have appeared over the last 15 years and from ongoing research (see, in particular, \cite{adls,aams,dr,bh,ahs1,ahs2,sloan,mercatti,nvm,aps1,aps3,asrev,iapsrev}.) While there is a uniform underlying theme, it is difficult for non-experts to appreciate that these results are synergistic and complement each other because the results are spread quite a bit in the literature and  presented using different notation and emphasis. We hope that this brief, unifying survey will  provide all researchers in gravitational physics with a broad overview of the current status.

We will focus on space-like singularities --such as the big bang and the big crunch that Paddy referred to, and the one inside the horizon of the Schwarzschild solution-- as they are generally considered to be places where space-time ends and physics comes to a halt. The abundance of such singularities in physically motivated solutions to Einstein's equations is brought to forefront by the celebrated singularity theorems of Penrose, Hawking, Geroch and others (see, e.g., \cite{he-book}). The key notion underlying these theorems is geodesic incompleteness. Physically, this corresponds to the property that trajectories of test particles come to an abrupt end and the tidal forces between them diverge. This then implies that the description of the physical world provided by general relativity using (pseudo-)Riemannian geometry fails at these singularities. The question is whether this is a genuine physical impasse or only a reflection of the inadequacy of notions normally used in general relativity. One can investigate this issue at several levels, using more and more sophisticated tools to incorporate an increasing number of features of the physical world, thereby making the discussion more and more reliable. We will discuss three such levels that provide new insights from different perspectives, each suggesting that these singularities are not as drastic as the standard framework of general relativity suggests.

In the first, one adopts the view that even at the classical level, Riemannian geometry of space-time should be regarded as `emergent' or `secondary notion' that arises in `tame' circumstances. While it is tremendously powerful when available, perhaps there is a more `fundamental' mathematical framework in terms of new variables from which space-time metric can be constructed. Dynamics of these variables may remain perfectly well-defined at space-like singularities where the Riemannian framework breaks down. If so, one would switch to the new  variables close to the singularity where Riemannian geometry is still well-defined, use the `fundamental framework' to evolve the system across the singularity, and then reintroduce the metric variables once we are `on the other side'. Of course the metric description will continue to fail at the singularity, but the premise is that the `more fundamental' description would not. In Section 2 we provide a concrete illustration of this possibility. Interestingly, the new description will not carry fields with space-time/manifold indices; the new variables will be (density weighted) space-time scalars that carry only `internal' SU(2) indices \cite{ahs1,ahs2}. Explicit calculations have been carried out in simple systems that exhibit the most interesting space-like singularities. They show that the `more fundamental' description does not break down. The dynamical equations for the new variables are available also for full general relativity, without any symmetry reduction. However, so far equations and their properties have not been analyzed in any detail for the full theory. Thus, level 1 should only be regarded as raising an unforeseen possibility rather than providing conclusive arguments.

Indeed, even in the simple examples, it is still true that the tidal forces between test particles become infinite there because curvature of the space-time metric diverges. Is this not a fundamental limitation of the framework? In our view this line of reasoning has a fundamental limitation. For, it is physically inappropriate to use \emph{classical} test particles as probes once the curvature becomes Planckian. Presumably, in this domain one has to use full quantum gravity. Nonetheless, already in the context of a classical space-time one can ask whether the singularity is a `death sentence' also to \emph{quantum} probes. Now, near space-like singularities, geometry is dynamical (not only in cosmology but also near the Schwarzschild singularity). Therefore one cannot meaningfully use single quantum particles as probes. Instead, probes have to be quantum fields. The question then is: Can we evolve quantum fields in the larger, extended space-time, e.g. of Section 2, across the singularity of the Riemannian geometry? To phrase this question precisely let us first recall that, already in Minkowski space-time, quantum fields $\hphi(x)$ are not operators but \emph{operator valued distributions} (OVDs). Therefore, we are led to ask if $\hphi(x)$ continues to be well-defined as an OVD even across the singularity? Again, rather surprisingly, the answer turns out to be in the affirmative. Encouraged by this result, one can ask more ambitious questions: Do the bi-distributions $\biphi$ continue to be well-defined? And what about the renormalized operator products $\phisq$ and $\stress$? Do they also continue to be well-defined as distributions? Answers turn out to be in the affirmative in spatially flat FLRW universes \cite{adls}, open and closed FLRW universes \cite{dr}. This tame behavior does not owe its origin to conformal flatness of FLRW space-times. In particular the field $\hPhi(x)$ need not be conformally coupled. It also extends to the Schwarzschild singularity where the Weyl curvature diverges \cite{aams}. Thus, when probed with quantum fields these singularities are tame. We will summarize these results in Section 3.

These results suggest that the menace posed by space-like singularities in standard, classical general relativity may be illusive. However, the framework used in Section \ref{s3} is  unsatisfactory for a fundamental reason: it uses a hybrid paradigm where geometry is treated classically and probes quantum mechanically. A full, self-consistent theory has to treat both quantum mechanically, and allow them to interact in a consistent manner. Since matter fields are OVDs, and expectation values such as $\stress$ are distributions, this suggests that quantum geometry will also have a distributional character. While we do not have a complete, fully satisfactory quantum gravity theory, loop quantum gravity already provides a detailed realization of this idea because its underlying quantum geometry is distributional in a precise sense \cite{alrev,crbook,ttbook,crfvbook,30years}. Thanks to this distributional nature, fundamental geometrical operators have discrete eigenvalues. In particular, there is an area gap $\Delta$ --the lowest non-zero eigenvalue of the area operator. As we explain in Section 4, in cosmological models, there is an upper bound on matter density and curvature, proportional to $(\Delta)^{\!-3}$. Therefore, there is no singularity. In place of a big-bang, there is a quantum bounce. The bounce has been studied in detail in loop quantum cosmology using a large number of models (summarized in \cite{asrev,iapsrev}), whence results on a natural singularity resolution due to quantum geometry are robust. 

Thus, at all three levels, the most important space-like singularities of classical general relativity appear to be harmless. Moreover, in these models, the three levels appear to be intertwined synergistically. However, as we discuss in Section 5, there is much room for further exploration to obtain a systematic understanding.

\section{Level 1: Reformulation of Classical General Relativity}
\label{s2}

While the powerful singularity theorems bring out the fact that it is rather common for the
gravitational field to develop singularities in general relativity, they provide little insight into the \emph{nature} of these singularities. The BKL conjecture \cite{bkl1} provides a key step in  filling this important gap. It posits that, as one approaches space-like singularities, time derivatives dominate over spatial derivatives, implying that the asymptotic dynamics would be well described by an ordinary differential equation. While the conjecture seems very surprising at first, by now there is considerable evidence in its favor \cite{berger,dg1,wib,ar,dg2,sag}. These investigations have shown that we can generally ignore matter in the analysis of the BKL behavior --the only matter that matters is a massless scalar field (or, phenomenologically, a stiff fluid). Therefore, without a great loss of generality one can restrict oneself to the gravitational sector of general relativity.%
\footnote{The interesting case of a massless scalar field is discussed in \cite{nvm} for FLRW cosmologies using the framework summarized here, and in \cite{sloan} for Bianchi IX models, using a technically different approach that also leads to a well-defined evolution through the big-bang.}
The reformulation summarized in this section provides precise statements of the (weak and strong forms of the) BKL conjecture in terms of variables that could be taken over to quantum theory more directly \cite{ahs1,ahs2}. In this section, however, we will focus only on the classical aspects.

Consider 4-dimensional space-times $(M, g_{ab})$ with $M= \M\times \mathbb{R}$. For simplicity of presentation, in this section we assume that the 3-manifold $\M$ is compact without boundary.  Let us first consider a triad version \cite{aa-newvar,alrev,ttbook} of the Arnowitt, Deser, Misner (ADM) 3+1 formulation of general relativity \cite{adm}. Thus, our canonically conjugate gravitational variables will be pairs $(E^a_i, K_a^i)$ of fields on $\M$, where $a,b,c \ldots$ are tensor indices and $i,j,k \ldots$ are ${\rm SO(3)}$ internal indices (which
can be freely raised and lowered using the Cartan Killing metric on ${\rm so(3)}$). $E^a_i$ represents an orthonormal triad on $\M$ \emph{with density weight 1} that determines a positive definite metric $q_{ab}$ on $\M$ via $E^a_i E^{bi}= q\, q^{ab}$, where $q$ is the determinant of $q_{ab}$. Similarly, on solutions, $K_a^i$ determines the extrinsic curvature of $\M$ via $K_{ab} = (1/\sqrt{q})\, q_{c(b}\, E^{c}_i K_{a)}^i$. The scalar, vector and Gauss constraints of vacuum general relativity are given by
\be S:= -q\,R - 2 E^a_{[i}E^b_{j]}\, K_a^i K_b^j = 0, \quad
V_a:= 4\, D_{[a}\,(K_{b]}^iE^b_I) = 0, \quad G^k
:=\epsilon_{i}{}^{jk} E^a_{j} K_a{}^i = 0\, ,\ee
where $D$ and $R$ denote the derivative operator and the Ricci scalar of $q_{ab}$. The Gauss constraint --which is absent in the ADM framework-- arises because there is more information in triads than in the metric; it simply ensures that the ${\rm SO(3)}$ triad rotations are gauge transformations. As in any background independent theory, the Hamiltonian generating dynamics is a linear combination of these constraints. 

The triad $E^a_i$ determines a unique ${\rm SO(3)}$ (or spin-)connection $\Gamma_a^i$ through $D_a E^b_i + \epsilon_{ij}{}^k \Gamma_a^j E^b_k = 0$. (Recall that $D$ is determined by $q_{ab}$; it ignores internal indices, treating fields with only internal indices as scalars.) For our purposes, it is more convenient to use all three fields, $\Gamma_a^i, K_a^i, E^a_i$, keeping in mind that $\Gamma_a^i$ is determined by $E^a_i$. The key idea behind the reformulation of general relativity equations is then the following. We first recall that at the big bang, big crunch and the Schwarzschild singularities the metric $q_{ab}$ becomes degenerate. Since its determinant $q$ vanishes there, one might expect that fields that are rescaled by appropriate powers of $q$ would remain well behaved at the singularity. Similarly, while the covariant spatial derivatives $D_af$ of a field $f$ may diverge at the singularity, derivatives $D_i f:= E^a_iD_a f$ could well be regular because of the $\sqrt{q}$ factor in the density weighted triad $E^a_i$.\,%
\footnote{This strategy is similar to that used in discussions of the BKL conjecture where one divides geometric fields by the trace $K$ of the extrinsic curvature which is expected to diverge at the singularity to obtain the so-called ``Hubble normalized fields" (see, e.g., \cite{UEWE}). A key difference is that whereas the focus in those treatments is on differential equations, here the focus is on a Hamiltonian framework that can serve as a point of departure for quantum theory.}
These motivations lead one to regard the following fields \emph{with only internal indices}
\be C_i{}^j:= E^a_i \Gamma_a^j - E^a_k \Gamma_a^k\, \delta_i^j ,
\quad {\rm and} \quad P_i{}^j:= E^a_iK_a^j - E^a_kK_a^k\,
\delta_i^j \, \ee
as basic variables. When $q_{ab}$ is invertible, one can freely pass between $(C_i{}^j,\, P_i{}^j)$ and $(\Gamma_a^i,\,K_a^i)$. However, because of the taming factor $\sqrt{q}$ in each of $E^a_i$, 
$C_i{}^j,$ and\, $P_i{}^j$ could be well defined at the singularity even when $\Gamma_a^i,\,K_a^i$ diverge. Interestingly, constraints can be re-expressed \emph{entirely} in terms of these new variables and their $D_i$ derivatives\,%
\footnote{Details on results summarized below can be found in \cite{ahs2}. The operator $D_i := E^a_i D_a$ is linear and satisfies the Leibnitz rule. Since $D_a$ is determined by $q_{ab}$, it ignores internal indices. Hence $D_i$ also treats fields with internal indices as scalars. Note however that, given a function $f$ on $\M$, $D_i f$  does not yield the exterior derivative $df$. Thus, $D_i$ is not a connection on $\M$. If we were to formally treat as a connection, it would have torsion determined by $C_i{}^j$: $D_{[i} D_{j]} f = -{T}^k_{\;\;ij} D_k f $ where ${T}^k_{\;\;ij} = \epsilon_{kl[i} C_{j]}^{\;\;l}$ \cite{bh,ahs2}.}
:
\ba \label{S1} S &:=& 2\epsilon^{ijk} D_i (C_{jk}) + 4 C_{[ij]}
C^{[ij]} + C_{ij}
C^{ji} - \f{1}{2} C^2 + P_{ij} P^{ji} - \f{1}{2} P^2 \approx 0\nonumber \\
\label{V1} V_i &:=& - 2D_j P_{i}{}^j - 2 \epsilon_{jkl}
P_i{}^j\, C^{kl} + \epsilon_{ijk} (2P^{jl}C_l{}^k - C P^{jk}) \approx 0\nonumber\\
\label{G1} G^{k} &:=& \epsilon^{ijk} P_{ji} \approx 0 \, .\ea
Note that the right sides contain low order polynomials of (density weighted) space-time scalars; tensor indices never appear. Furthermore, $(C_i{}^j,\, P_i{}^j)$ are closed under Poisson brackets:  
\ba \label{pb2} \{ \textstyle{\int} f_{ij}(x) P^{ij}(x),\, \textstyle{\int} g_{kl}(y) P^{kl}(y) \} &=& \textstyle{\int} \big(f_{kj}(x) g_i^k (x) - f_{ik} g^k{}_j\big) P^{ij}\nonumber\\
\label{pb3} \{ {\textstyle{\int}} f_{ij}(x) P^{ij}(x),
{\textstyle{\int}} g_{kl}(y) C^{kl}(y) \} &=& {\textstyle{\int}}
\big( f_{ij} g_{kl} (C^{kj} \delta^{il} + C^{jl} \delta^{ik})
+\epsilon^{jlm} \delta^{ik} g_{kl} D_m f_{ij}\big)\nonumber\\
\label{pb4} 
\quad \{\textstyle{\int} f_{ij} C^{ij}(x),\, \textstyle{\int} g_{lk}(y) C^{kl}(y)\} &=&0\, , \ea
where $f_{ij}(x)$ and $g_{kl}(x)$ are smooth test scalar fields. The Poisson brackets between $E^a_i$ and $C^{ij},\, P^{ij}$ are also simple:
\be \label{pb1} \{ \textstyle{\int}\omega_a^i E^a_i,\, \textstyle{\int} f_{jk}\, P^{jk} \} =
\int \big(\omega_a^k f^j{}_k E^{a}_j - \omega_a^i f_j{}^j E^a_i\big); \quad 
\{ \textstyle{\int}\omega_a^i E^a_i,\, \textstyle{\int} f_{jk}\, C^{jk} \} =0\, , \ee
where $\omega_a^i$ is a Lie-algebra valued, smooth, test 1-form. Since the dynamics is generated by constraints, evolution equations of $C^{ij}$ and $P^{ij}$ involve \emph{at most quadratic} combinations of these variables and their $D_i$ derivatives. Given any scalar density $s_{(n)}$ of weight 1, one can obtain the evolution equation of $D_i s_{(n)}$ using (\ref{pb1}) and the right side involves only a linear term in $P_i^j$ and $D_i P_j{}^j$.%
\footnote{ By contrast, evolution equations of the ADM variables $(q_{ab}, P^{ab})$ as well as their triad analogs $(E^a_i, K_a^i)$  involve non-polynomial functions of these variables and, furthermore $P^{ab}$ and $K_a^i$ themselves diverge at the FLRW big bang.} 
Thanks to this simplicity, one can hope that the evolution of $C^{ij}$ and $P^{ij}$ would be well-defined across the singularity.

As indicated in Section \ref{s1}, these equations are to be used as follows. Consider a Cauchy surface $\mathbb{M}$ near the singularity where $q_{ab}$ is well defined and non-degenerate. Then we can construct $(C_{ij}, P_{ij}, D_i)$ from a pair $(E^a_i, K_a^i)$ of canonical variables. We can then work exclusively with the triplet $(C_{ij}, P_{ij}, D_i)$. On $\mathbb{M}$, the pair $(E^a_i, K_a^i)$ satisfies constraints if and only if the triplet satisfies (\ref{S1})--(\ref{pb4}). Given such a triplet, we can evolve it using their Poisson brackets with constraints, \emph{without having to refer back to} $E^a_i$. The end product of this evolution would provide us with a 1-parameter family of fields $C_{ij}(t), P_{ij}(t)$ and the operator $D_i(t)$ with only internal indices. However, to recover  the familiar description in terms of Riemannian geometry, one does need the density weighted triad. One can calculate the Poisson bracket between $E^a_i$ and the constraints using (\ref{pb1}) to obtain the time derivative of $E^a_i$ in terms of $C_{ij}(t), P_{ij}(t)$ and $E^a_i$ itself. Given  $C_{ij}(t), P_{ij}(t)$, one should be able to solve this ordinary differential equation to obtain $E^a_i$ at the singularity and evolve it further.
 
So far this program has been carried out \cite{nvm} only in models with a high degree of symmetry: (i) Friedmann cosmologies with a massless scalar field as source. In this case matter \emph{does} matter, affecting, for example, the BKL behavior \cite{ar} because the blow up of matter density is much more severe than in other commonly used sources; (ii) Bianchi I and Bianchi IX models; and (iii) The portion of the Schwarzschild space-time inside the horizon. In all these cases, the basic fields $C_{ij}, P_{ij}$ remain well-defined at the singularity and, as expected, the density weighted triad $E^a_i$ vanishes there. Therefore one can evolve across the singularity. From the viewpoint of differential equations, this is not surprising because, as pointed out in \cite{dacr}, a singularity can arise if equations are written in terms of inconvenient variables as in the following trivial example: while a solution $y= 1/\sin t$ of $y\ddot{y} -\dot{y}^2 + y^2 =0$ is singular at $t=0$, in terms of $x= 1/y$  the equation becomes $\ddot{x} + x =0$ for which the solution $x(t)= \sin(x)$ can be continued across $x=0$ smoothly. The question of course is which variables are the physical ones. Even if one were to say that $y$ is the physical variable, passage to $x$ provides a trivially streamlined procedure to continue the solution beyond $t=0$. In our examples, the situation is better in that not only are the new variables $C_{ij}, P^{ij}$ and $E^a_i$ well-defined at the singularity but the 3-metric $q_{ab}$ is also well-defined as a tensor field. From the perspective of Riemannian geometry, the problem is that $q_{ab}$ becomes degenerate at the singularity causing the space-time curvature to diverge. As discussed in Section \ref{s1}, the question is whether these divergences imply that physics breaks down rather than just the classical description in terms of Riemannian geometry. Regularity of equations in terms of $C_{ij}, P_{ij}, D_i$ in physically interesting models is an indication that it may be the latter.

\section{Level 2: Using Quantum Fields as Probes}
\label{s3}

To begin with, let us focus on the big-bang of cosmological models and probe the singularity with test \emph{fields} rather than test-particles. Because FLRW models are conformally flat, the conformally \emph{invariant} classical Maxwell field $F_{ab}$ remains regular at the big-bang. The situation is different for scalar fields. When evolved back in time, solutions $\phi(x)$ to the minimally coupled scalar field equations, for example, generically diverge at the big bang. (The situation is the same also for conformally coupled scalar fields because the equation they satisfy is only conformally covariant; not conformally invariant.) Now, in physical applications of the classical theory one expects $\phi(x)$ to be a suitably smooth function that obeys its field equation in the corresponding smooth sense. A blow-up of $\phi(x)$ signals a blow-up of physical observables such as energy density. In quantum theory, one generally expands the quantum analog $\hphi(x)$ of $\phi(x)$ as a sum of creation and annihilation operators with a basis of positive and negative frequency classical solutions as coefficients. Since these solutions diverge at the big-bang, one's first reaction may be that the quantum field $\hphi(x)$ is also ill-defined there.

 Note, however, that there is a key conceptual and mathematical difference between a classical field $\phi(x)$ and its quantum analog $\hphi(x)$. Already in Minkowski space, $\hphio(x)$ is an operator valued \emph{distribution} (OVD): we have to smear it with test fields $f(x)$ to obtain well-defined operators $\phio(f) := \int_{\mathring{M}} \phio(x) f(x) \rmd^4 x$ on the Fock space \cite{asw}. (Throughout, fields in Minkowski space-time will carry a marker $\circ$.) The distributional character is not a mere technicality; it lies at the foundation of quantum field theory. In particular, for $\hphio(x)$ satisfying the wave equation with respect to the Minkowski metric 
 $$\mathring{g}_{ab}\rmd x^a \rmd x^b \equiv \rmd \mathring{s}^2 = -\rmd \eta^2 + \rmd \vec{x}^{\,2},$$ we have:
\ba 
 [\hphio(x), \, \hphio(\xp)\, ] &=& i\hbar\, (G_{\rm ad} - G_{\rm ret})(x,\, \xp)\, \h{I} \quad  {\rm and} \nonumber \\ 
\biphio  &=&  \f{\hbar}{4\pi^2} \, \f{1}{|\vx-\vxp|^2 - ((\eta-\etap)-i\epsilon)^2} \ea
where the right sides are genuine distributions. In the second equation one has to first integrate $\biphio$ with $i\epsilon$ against test functions and \emph{then} take the limit $\epsilon\to 0$. More importantly, products $(\hphio){}^2(x)$ have to be regularized precisely because $\hphio(x)$ is an OVD. Thus, if taken literally, the textbook terminology of `field operators' and `2-point functions' can be quite misleading, just as the term Dirac `$\delta$ function' is. 

Since $\hphi(x)$ is an OVD already in Minkowski space-time, we cannot hope for a smoother behavior in FLRW space-times. Thus, we are led to ask: do the operators $\hphi(f) := \int \rmd^4V \hphi(x) f(x)$, smeared with test functions $f(x)$, remain well-defined and continue to satisfy the field equation $\Box\, \hphi(x) =0$ in the distributional sense (i.e., $\int \rmd^4 V\,\hphi(x)\, (\Box f(x)) =0$), even when the support of the test field $f(x)$ \emph{includes} the singularity? As usual, here $\rmd^4 V$ is the space-time volume element determined by the space-time  metric $g_{ab}$ of the FLRW space-time. Now, in these space-times, the volume element shrinks at the big bang because the scale factor vanishes there. Therefore, it is `easier' for $\h\phi(x)$ (as well as the classical solutions $\phi(x)$) to be  well-defined as distributions. A simple analogy is provided by $h(\vec{r}) := 1/r$ on $\mathbb{R}^3$: it is singular as a function but well-defined as a distribution. Indeed, we routinely use the equation $\vec\nabla^2\, h(\vec{r}) = - 4\pi\, \delta(\vec{r})$ in the distributional sense. 

It turns out that the divergence of the basis functions is exactly compensated by the vanishing of the volume element, making $\hphi(x)$ a well-defined OVD in spite of the big-bang singularity. One can see a shadow of this phenomenon already in the classical theory. Consider the `symplectic product' $\Omega(\phi_1, \phi_2) := \int_{M_\circ} [\phi_1 (x) \nabla_a \phi_2(x) - \phi_1 (x) \nabla_a \phi_2(x)]\, n^a\, \rmd^3 V$ between any two solutions, where $M_\circ$ is a Cauchy surface and $n^a$ is the unit normal to it. Since $\phi_1(x)$ and $\phi_2(x)$ both diverge at the singularity, one might first think that the right side would diverge as we push $\mathring{M}$ to the big-bang. However, the symplectic product is conserved, i.e., is independent of the choice of $\mathring{M}$. Therefore we can take $\mathring{M}$ to be a $t={\rm const}$ surface, where $t$ is the proper-time of the FLRW metric, and take limit as $t\to 0$, i.e.,  approach the big bang. The limit of $\Omega(\phi_1, \phi_2)$ is obviously well-defined (since it remains constant in the process) although $\phi_1(x)$ and $\phi_2(x)$ diverge. This is because the divergence is exactly compensated by the shrinking of the volume element.

For brevity, let us focus on spatially flat FLRW space-times. As is well-known the analysis of  quantum fields simplifies if one uses conformal time $\eta$. With this choice, the metric has a manifestly conformally flat form: $g_{ab} \rmd x^a \rmd x^b = a^2(\eta) \mathring{g}_{ab} \rmd x^a \rmd x^b \equiv a^2(\eta)(-\rmd\eta^2\,+\,\rmd \vec{x}^{\,2})$, where $a(\eta) = a_\beta \eta^\beta$ with $a_\beta$ a constant and $\beta \ge 0$. The case $\beta=0$ corresponds, of course, to Minkowski space-time for which $\eta$ runs over the full real line of the Minkowski 4-manifold $\mathring{M}$. Next, $\beta=1$ corresponds to the radiation filled universe and $\beta=2$ to the dust filled universe. In these cases, we have a big bang at $\eta=0$ and physical space-time corresponds only to the portion $\eta>0$ of Minkowski space-time $(\mathring{M}, \mathring{g}_{ab})$. The question is: Can the evolution of the quantum field $\hphi(x)$ be nonetheless extended to all of $\mathring{M}$? To explain the main ideas, we will restrict ourselves to the simplest non-trivial case of a radiation-filled universe. Treatment of higher values of $\beta$ is slightly more complicated because of a technical infrared issue that are  unrelated to the big-bang singularity. We will only indicate these complications and refer the reader to \cite{adls} for details. Qualitative features of the behavior of quantum fields are the same for higher values of $\beta$.

To analyze the evolution across the big-bang, we need to extend the metric $g_{ab}$ to full $\mathring{M}$, i.e., to $\eta \le 0$. We will achieve this simply by extending the conformal factor $a^2(\eta)$ in the obvious manner. In the radiation-filled universe, then, $a^2(\eta)= a_{1}^2\,\eta^2$ for all $\eta \in \mathbb{R}$ and $g_{ab}$ is in fact smooth as a tensor field on all of $\mathring{M}$. (This is also the extension provided by the framework of Section \ref{s1}.) But $g_{ab}$ fails to be invertible at $\eta=0$, whence its curvature diverges there. The key issue is whether our test quantum fields experience this divergence and, if they do, whether it is as fatal for them as it is for classical test particles. 

This issue has been analyzed in the general setting provided by tempered distributions \cite{schwartz,gs} where test functions are taken to lie in the Schwartz space $\S$ of $C^\infty$ functions that decay faster than any polynomial at infinity (in the Cartesian coordinates of Minkowski space $(\mathring{M}, \mathring{g}_{ab})$). $\S$ is a more convenient choice of test functions than the more commonly used space $C^\infty_0$ of smooth functions of compact support because, while $\S$ is stable under Fourier transforms, the space $C^\infty_0$ is not. Note that since $C^\infty_0 \subset \S$, a tempered distribution is in particular an ordinary distribution which acts on smooth functions of compact support. Therefore, if $\hphi(x)$ is well-defined as a tempered OVD, it is in particular a well-defined OVD over $C^\infty_0$. A key fact about tempered distributions is that, while the function $\eta^{-m}$ is divergent at $\eta=0$  on the real line, it is a well-defined tempered distribution \cite{schwartz,gs}, denoted by ${\underline{\eta}}^{-m}$:
\be \label{xminusm} {\underline{\eta}}^{-m} := \f{{(-1)}^{m-1}}{(m-1)!}\,\f{\rmd^m \ln |\eta|}{\rmd \eta^m};\quad {\rm i.e.},\quad
\underline{\eta}^{-m}:\,\, f(\eta) \quad\rightarrow\quad  \f{{(-1)}^{m}}{(m-1)!}\, \int\!\rmd \eta\, \ln |\eta|\,\,\f{\rmd^m f}{\rmd \eta^m} \,\,  \ee
for all $f \in \S$. This definition of ${\underline{\eta}}^{-m}$ is completely analogous to the definition of the more familiar distribution $\delta^m(x)$ --the m-th derivative of the Dirac distribution-- which is defined as the $(m+1)$th derivative of the locally integrable but non-differentiable step function.

With these notions at hand, we can now summarize the results \cite{adls}. Let us consider the OVD $\hphi(x)$ in radiation-filled FLRW space-time $(M, g_{ab})$ satisfying the wave equation $\Box \hphi(x)=0$ (for $\eta >0$). Then, $\hphio(x):= (a_1\eta)\hphi(x)$ satisfies  $\mathring{\Box}\, \hphio(x)=0$ with respect to the Minkowski metric $\mathring{g}_{ab}$. Therefore, we can use the Minkowskian expansion of $\hphio(x)$ in terms of its creation and annihilation operators (which is well-defined on full $\mathring{M}$) and obtain a putative OVD $\hphi(x) := (a_1\eta)^{-1}\, \hphio(x)$ on $\M$. Is it a well-defined OVD on our extended FLRW space-time on which $g_{ab}$ is degenerate at $\eta=0$ (and has a curvature singularity there)? Using the fact that the volume elements of the two metrics are related by $\rmd^4 V = (a_1 \eta)^4\, \rmd^4 x$, we obtain  
\be \h\phi(f) := \int_{\mathring{M}} \rmd^4 V \hphi(x) f(x) = \int_{\mathring{M}} d^4x\, \hphio(x) ((a_1 \eta)^3 f(x))\ee
and the right side is a well-defined operator on the Minkowski Fock space  since $(\eta^3 f(x))$
is in the Schwartz space $\S$ if $f$ is. Since $f$ can have support at $\eta \le0$, $\hphi(x)$ is a well-defined OVD on the full extended FLRW space-time; its explicit action can be given in terms of the creation and annihilation operators of the Minkowski Fock space. Similar argument shows that $\hphi(x)$ satisfies $\int_{\mathring{M}} \rmd^4V\, \hphi(x) (\Box f(x)) =0$ even when $f$ has support on $\eta \le 0$; thus our OVD satisfies the wave equation on the extended space-time. For the bi-distribution $\biphi$ we obtain: $\biphi  = \f{1}{a_1^2 \eta \eta^\prime} \, \langle \hphio(x)\, \hphio(\xp)\rangle_{{}_{\circ}}\,$ which is a well-defined bi-distribution on the extended FLRW space-time, again because the double volume element goes as $(a_1^2 \eta\eta^\prime)^4$. By considering the \emph{physical} geodesic distance between two points one can check that the singularity structure has the Hadamard form. (For a discussion of the Hadamard form, see, e.g. \cite{rmw-book,cfkr}.)

Finally, one can consider renormalized products of OVDs $\phisq$ and $\stress$. The renormalization procedure involves subtracting a counter term that removes the `universal' part of the divergence. The standard strategy is to first carry out point splitting \cite{bsd1,bsd2,smc1,smc2} and then subtract the DeWitt-Schwinger subtraction term \cite{bsd2}, constructed from curvature tensors of the background space-time metric. Applying this procedure to the radiation filled universe, one obtains  $\phisq = 0$ and  
\be\stress \,=\, \frac{\hbar} {720\pi^2 a_1^2\eta^6}\, \nabla_a\eta \nabla_b\eta + \frac{\hbar}{576\pi^2 a_1^2\eta^6}\, \go_{ab}\, \equiv\, T_1(\eta)\, \nabla_a\eta \nabla_b\eta \,+\, T_2(\eta)\, \go_{ab}\, .\ee
The first result seems surprising because $\phisq$ is generically non-zero in FLRW space-times. However, dimensional considerations show that it can only be proportional to the space-time scalar curvature which happens to vanish in the radiation-filled FLRW cosmologies.  (In the dust-filled case, for example, it is non-zero but a well-defined distribution.) The coefficients $T_1(\eta)$ and $T_2(\eta)$ in the expression of the renormalized stress-energy tensor diverge as $1/\eta^6$. Therefore, when the volume element is folded in, they become tempered distributions of the type $1/\underline{\eta}^2$. Thus even the renormalized operator products are well-defined as distributions.

As we just saw, technical simplifications arise in radiation filled universes, because the scalar curvature $R$ of the space-time metric $g_{ab}$ vanishes identically. In particular, the natural mode functions (in the expansion of $\hphi(x)$ in terms of creation and annihilation operators) in the FLRW space-time can be obtained simply by rescaling the Minkowski mode functions\, $e^{-ik\eta + i \vec{k}\cdot\vec{x}}/\sqrt{2k}$\,\, by\,\, $1/a(\eta) = 1/(a_1 \eta)$. For FLRW universes with scale factor $a(\eta) = a_\beta \eta^\beta$ with $\beta >1$, the scalar curvature does not vanish. Nonetheless, a simplification occurs because $\Box \phi(x) = 0$ if and only if $(\mathring\Box + \beta(\beta -1)/\eta^2) \phio(x) =0$, where $\phio(x) = a(\eta)\phi (x)$. Therefore one can focus on $\phio(x)$ satisfying the wave equation in presence of a time dependent potential\,\, $V(\eta) = \beta(\beta -1)/\eta^2$. Because the potential decays sufficiently rapidly at infinity and is symmetric around $\eta=0$, one can again find canonical mode functions --or, more precisely, a complex structure on the space of solutions (see, e.g., \cite{am}), or, a quasi-free vacuum state (see, e.g., \cite{cfkr}). For $\beta >1$, the explicit form of the corresponding mode functions is more complicated, but one can still carry out the calculations we reported above for radiation-filled universes. 

Detailed calculations are available in dust-filled universes ($\beta=2$). The only significant complication is that one has to introduce an infrared regularization procedure (also for $\beta > 2$). But this issue has nothing to do with the big-bang; it arises already in the usual portion of Friedmann universes where we have $\eta >0$ \cite{fp}. Once an infrared regulator is introduced, the theory is well-defined for $\eta>0$ and then there is no further difficulty in extending it to the full manifold $\mathring{M}$. This is just as one would expect, given that possible difficulties at the big-bang are related to the ultraviolet regime rather than the infrared one. $\hphi(x)$ is again a well-defined OVD and $\biphi$ a bi-distribution on all of $\mathring{M}$. Both $\phisq$ and $\stress$ continue to be well-defined distributions but now $\phisq$ is non-zero. While our discussion is restricted to spatially flat FLRW models, recently these results have been extended to include spatial curvature as well \cite{dr}. Interestingly, in dust-filled universes which have been analyzed in detail, the infrared issue mentioned above disappears because the presence of spatial curvature introduces a new scale which in effect provides a natural infrared cutoff.

To summarize, then, the big-bang and the big-crunch singularities of general relativity are harmless when probed with quantum fields. The divergence of the curvature does have direct consequences. Conceptually $\phisq$ and $\stress$ should be regarded as distributions --indeed $\biphi$ is already a bi-distribution-- but away from the singularity these distributions happen to be functions, taking well-defined values at points of $M$. On the extended manifold $\mathring{M}$, they become genuine distributions of the type $\underline{\eta}^{-m}$ --reminiscent of $1/r$ on $\mathbb{R}^3$. Finally, the space-time region inside the Schwarzschild horizon is time-dependent and isometric with the Kantowski-Sachs cosmological solution which is spatially homogeneous but not isotropic. Quantum fields have also been analyzed in this space-time \cite{aams} and results to date again show that the singularity is harmless when probed with quantum fields.%
\footnote{Sometime ago a similar result was obtained using Schr\"odinger picture for time evolution \cite{hs}. However, that reasoning relies on formal arguments that do not do full justice to the difficult quantum field theoretic issues associated with an infinite number of degrees of freedom. These are adequately handled in \cite{aams}, along the lines of \cite{adls}.}

\section{Level 3: Using Quantum Riemannian Geometry}
\label{s4}

Up to this point we have treated the gravitational field classically. As mentioned in Section \ref{s1}, a more complete theory would treat both matter \emph{and} gravitational fields quantum mechanically. Thus at a fundamental level, space-time geometry itself would have a quantum or `atomic' structure and the continuum picture in terms of a metric $g_{ab}$ would arise only on suitable coarse graining of the underlying `atoms of geometry'. The fact that the renormalized stress-energy tensor has to be treated as a genuine distribution near the singularity suggests that, when probed at the Planck, geometry would also display a `distributional character'. 

In loop quantum gravity (LQG) this idea is realized in a concrete fashion. Fundamental excitations of geometry are 1-dimensional, rather like polymers that serve as `threads' with which the quantum geometry is `woven'. The basis of states that diagonalizes the geometric operators is given by spin networks --graphs whose links and nodes carry certain quantum labels. Quanta of volume reside only at the nodes; the label carried by a node tells us which quantum that reside there. Each link deposits a certain quantum of area on any surface it intersects; the label carried by the link  determines the eigenvalue of the area operator deposited. In this precise sense, area and volume are distributional in character. In particular, given a spin network state of quantum geometry, an open region that has no nodes has zero volume, and a portion of a surface that has no intersections with any of the links has zero area. If a region or a surface is  macroscopic and the state is semi-classical with many, many nodes and links, then this fundamentally discrete quantum geometry can be well-approximated by a continuum. This is analogous to the fact that while a coherent state peaked at a large classical electromagnetic field $F_{ab}$ is fundamentally a superposition of photon states, one can approximate it by the classical $F_{ab}$ extremely well. 

Therefore it is interesting to examine classical singularities from the perspective of the quantum Riemannian geometry. Does the fundamental atomic structure of geometry come to rescue and naturally resolve the singularity? There is very extensive work on this subject in loop quantum cosmology (LQC). (For reviews, see, e.g., \cite{asrev,iapsrev}.) In this section, we will summarize the main ideas and results on the fate of cosmological singularities in LQC. 
 
 The classical FLRW spacetime is characterized by  a scale factor $a(t)$ together with matter fields, say $\phi(t)$. In any quantum cosmology, the classical space-time is replaced by a quantum state $\Psi(a,\phi)$ that is subject to the \emph{quantum versions} of the Friedmann and Raychaudhuri equations. In this transition, reference to the proper time $t$ disappears --we only have $a$ and $\phi$ and quantum dynamics becomes `relational'. One can use, for example, the matter field $\phi$ as an internal clock, and describe how the scale factor evolves with respect to it. Quantum cosmology was first introduced in the ADM framework and the quantum version of the ADM scalar/Hamiltonian constraint goes under the name \emph{Wheeler-DeWitt} (WDW) \emph{equation.} Let us make a small detour to compare and contrast LQG with the WDW theory, also known as Quantum Geometrodynamics. For full general relativity, the mathematical framework underlying Quantum Geometrodynamics has remained formal; difficult issues of functional analysis remain unresolved even now. Even within quantum cosmology --where one only has a finite number of degrees of freedom--  issues such as the inner product on physical states --the space of solutions to the Wheeler DeWitt {\rm (WDW)} equation-- or self-adjointness of physical observables, or the measure used in the definition of path integrals is rarely discussed (outside the LQC community). LQG differs from the WDW theory in this respect. First, LQG is based on a rigorous mathematical framework in which functional analytic issues related to the presence of an infinite number of degrees of freedom are handled systematically respecting covariance, i.e., the fact that there are no background fields, not even a space-time metric. It is this feature that leads one to a specific quantum Riemannian geometry in which geometric operators have discrete eigenvalues \cite{almmt1,rs4,loll1,al5,al6,tt2,eb1}. As mentioned in Section \ref{s1}, the area operator has a minimum non-zero eigenvalue $\Delta \approx 5.17\ell_{\rm Pl}^2$, called the area-gap, which has no analog in Quantum Geometrodynamics. The area gap plays an important role in the theory because the curvature operator is defined by computing the holonomies (or Wilson loops) of the gravitational connection around closed loops and then shrinking the loop till area it encloses becomes $\Delta$. This  introduces a Planck scale non-locality in the theory that then provides a natural ultraviolet cut-off.

The mathematical framework of LQC descends in a precise sense from that of full LQG \cite{ac,eht}. As a result, already at the kinematical level--i.e. even before imposing the quantum constraint-- the  Hilbert space of states is not the space of square integrable functions of, say, $(a,\phi)$ with respect to the Lebesgue measure, the arguments of the wave function. One is led to a novel Hilbert space $\H$, even though the system has only a finite number of degrees of freedom \cite{aps1,aps3}. The resulting theory is sometimes referred to as ``polymer quantum mechanics'' because it descends from full LQG where, as we mentioned above, the fundamental excitations of geometry are polymer-like. Furthermore, the Wheeler DeWitt differential operator on $\Psi(a,\phi)$ --the quantum version of the Hamiltonian constraint in Quantum Geometrodynamics-- fails to be well-defined on the LQC $\H$ but gets systematically replaced by a \emph{difference operator}  that does have a well-defined action on $\H$. The step size of this difference operator is dictated by the area gap $\Delta$. As a consequence, quantum dynamics has qualitatively new features at the Planck scale.

Because there is a well-defined Hilbert space, given a state $\psi(a,\phi)$ one can calculate the expectation values and uncertainties of operators in it. Therefore there is a well-defined notion of ``sharply peaked states''. Fix a classical FLRW cosmology with $\phi$ as source and consider a dynamical trajectory $a(t), \phi(t)$. As we go back in time along the trajectory, the scale factor goes to zero and the energy density of the scalar field diverges. What happens in LQC? Let us take  a quantum state $\Psi(a, \phi)$ that is sharply peaked on the given classical trajectory at a late time, i.e., where  spacetime curvature and matter density are low compared to the Planck scale. Let us then use the LQC Hamiltonian constraint operator to evolve this state back in time (w.r.t. the internal `matter clock') towards higher curvature and density.  Interestingly, the wave packet remains peaked and the peak follows the classical trajectory till the density increases to about  $\rho\, \sim 10^{-4} \rho_{\rm Pl}$. Then the quantum geometry effects induced by the area gap --i.e., the finite size of steps in the difference operator representing the Hamiltonian constraint-- cease to be negligible. The state $\Psi(a,\phi)$ is still sharply peaked but the peak ceases to follow the classical trajectory. Rather, it follows a \emph{quantum corrected} trajectory that undergoes a bounce when the density reaches  a critical, maximum value 
\be \rho_{\rm max}  := \f{18\pi G\hbar^{2}}{{\Delta}^{3}} \, \approx 0.41 \rho_{\rm Pl}\ee
and then it starts decreasing. \emph{The bounce replaces the big-bang}. In this backward evolution, quantum corrections become negligible and GR is again an excellent approximation once the density falls to $\rho \sim 10^{-4} \rho_{\rm Pl}$  (see, e.g., \cite{aps1,aps3,asrev,iapsrev} for details of these results). Thus, quantum geometry effects create a bridge joining our expanding FLRW branch to a contracting FLRW branch in the past. This behavior is strikingly different from that of the WDW wave function $\Psi_{\rm WDW}(a,\phi)$ which evolves into a singularity; there is no bounce. What would happen if we ignored the quantum geometry effects? In dynamical equations, this would correspond to steadily decreasing the area gap $\Delta$ to zero. One arrives at the WDW evolution in a precise sense \cite{acs}. The big bang persists because $\lim_{\Delta \to 0}  \rho_{\rm max} = \infty$. Thus, the replacement of the big bang by a big bounce can be directly traced back to quantum geometry effects of LQG. These qualitative new features are consequences just of the quantum Einstein's equations given by LQC. One does not introduce matter violating energy conditions to escape singularity theorems, nor does one introduce new boundary conditions, such as the Hartle-Hawking `no-boundary proposal' \cite{hh}.

Salient features of the LQC dynamics can be understood using certain systematically derived  \emph{effective equations} that capture the qualitative behavior of sharply peaked states $\Psi(a,\phi)$ \cite{vt,asrev,iapsrev}. More precisely, they encode the leading order corrections to the classical Einstein's equation in the Planck regime. As we discussed above, these corrections modify the geometrical part (i.e. left side) of Einstein's equations due to quantum geometry effects. But it is both possible and convenient to move them to the right side by a simple mathematical manipulation. Then, the quantum corrected Friedmann equation assumes the form
\be \Big(\frac{\dot{a}}{a} \Big)^2 = \frac{8\pi G\,\rho} {3}\, \left(1 -  \frac{{\rho}}{\rho_{\rm max}} \right) \, . \ee
where the second term on the right side represents the quantum correction. Without this term, i.e., in classical GR,  the right side is positive definite, whence $\dot{a}$ cannot vanish at any finite time: the universe either expands out from the big bang or contracts into a big crunch. But, with the quantum correction, the right side vanishes at $\rho =\rho_{\rm max}$. Therefore $\dot{a}$ vanishes there and
the universe bounces. Note that this can occur only  because the LQC correction $\rho/\rho_{\rm max}$ \emph{naturally} comes with a \emph{negative} sign, which effectively gives rise to a `repulsive force' of quantum geometry origin in the Planck regime. The occurrence of this negative sign is non-trivial: in the standard brane-world scenario, for example, Friedmann equation is also receives a $\rho/\rho_{\rm max}$ correction, but it comes with a positive sign (unless one makes the brane tension negative by hand; see, e.g. \cite{sahni}). Therefore the singularity is not naturally resolved. Finally, there is an excellent match between analytical results within the quantum theory, numerical simulations and effective equations.  On the conceptual side, this singularity resolution first arose in a Hamiltonian framework \cite{mb1,aps1,aps3}. However, subsequently they were derived in the sum over histories approach as well \cite{ach}, and also understood using the `consistent  histories framework' of quantum mechanics \cite{dcps}. Finally, these considerations have been generalized to include spatial curvature, non-zero cosmological constant, and anisotropies (see, e.g., \cite{asrev,iapsrev} and references therein). The simplest inhomogeneities captured by the Gowdy models \cite{gowdy} have also been included in the analysis and Brans-Dicke theory \cite{bdtheory} has been discussed as well. Finally, we note that there is considerable work on evolving test quantum fields on the FLRW \emph{quantum} geometries $\Psi(a,\phi)$ of LQC, especially in the context of cosmological perturbation theory (\cite{akl}, see also e.g., \cite{abrev,iapsrev} and references therein).

\section{Outlook}
\label{s5}

It is interesting to recall Einstein's own perspective on the big-bang singularity in his later years. In the 1945 edition of his  \emph{Meaning of Relativity}, he summarized his evolved views as follows:
\begin{quote}
{\sl ``One may not assume the validity of field equations at very high density of field and matter and one may not conclude that the beginning of the expansion should be a singularity in the mathematical sense."}
\end{quote}
By ``field equations" he meant his own equations of general relativity and the suggestion that these equations would have to be supplanted or transcended at high field density (i.e. curvature) and high matter density is striking. 

Einstein's remarks are in the context of the big bang singularity. In the last three sections, we presented an overview of the current understanding of the broader question of whether space-like singularities of classical general relativity should be viewed as the `final boundaries' at which space-time ends and physics breaks down. We explored this issue at three different levels. The premise of the first-level investigation was that while the evolution of the space-time metric does break down, there may be other `more basic' classical variables which are not singular there. The premise of the second-level exploration was that while the space-time metric does become singular, this singularity may be tame when probed with test quantum fields. Once we give a precise {\sl `mathematical sense'} to these probes, taking into account their distributional nature, there is no singularity in their evolution. At the third level, the premise is that one {\sl `should not assume the validity of Einstein's field equations'} in the Planck regime where there is a very {\sl `high density of field and matter'.}  Interestingly, at each level, the analysis offered hints that the answer to the broader question is likely to be in the negative, echoing Einstein's sentiments. 

At the outset, the three levels are conceptually very different from each other. Nonetheless, there are common underlying threads. The variables $C_{i}{}^j$ and $P_{i}{}^j$ of Section \ref{s2}, for example, are real and imaginary parts of $(E^a_i A_a{}^j)$, where $A_a^i$ is the self-dual gravitational connection that plays a key role in the Hamiltonian formulation underlying loop quantum gravity, discussed in Section \ref{s4}. So the fact that evolution of  $C_{i}{}^j$ and $P_{i}{}^j$ is well defined across the big bang  may be a shadow on the classical theory of the fact that the evolution of the quantum state $\Psi(a,\phi)$ in LQC does not break down because of the big-bang singularity. Similarly, the tame behavior of test quantum fields across the big-bang we found in Section \ref{s3} may be a reflection of the fact that quantum fields can be meaningfully evolved on the singularity-free quantum geometries represented by wave functions $\Psi(a,\phi)$ of LQC. Put differently, it may well turn out that the tame behavior we found at levels 1 and 2 are systematic consequences of the regular evolution in a quantum theory of gravity, across the putative classical singularities. For example, it may be possible to \emph{derive} them step by step starting from LQG, perhaps after its full quantum dynamics is better understood. In particular, one could study in detail the Hamiltonian system offered by the variables $C_{i}{}^j$ and $P_{i}{}^j$ of Section \ref{s2} at the classical level, beyond the simple (but physically interesting) symmetry reduced versions that have been studied so far. One could also investigate quantization of that system. Similarly, one could explore the relation between the distributional fields --especially the renormalized stress-energy tensor-- we discussed in Section \ref{s3} and the distributional geometries underlying LQG we referred to in Section \ref{s4}. These are fascinating challenges for future work.

In this regard, a key question for a theory such as LQG can be phrased as follows. In classical general relativity, one first observed that physically important, explicit solutions to Einstein's equations such as homogeneous cosmologies and the Schwarzschild black holes have curvature singularities. But then there was a long debate as to whether these singularities should be taken seriously. Eddington famously claimed that a gravitational collapse of a star will not lead to a singularity; {\sl ``I think there must be a law of Nature to prevent the star from behaving in this absurd way''.} Later, there was a systematic program led by Khalatnikov and Lifshitz whose goal was to show that the singularities found in these known solutions were artifacts of their high symmetry and would be absent in a `general solution' to Einstein's equations. Singularity theorems bypassed the analytical methods that were then  being used to find `general solutions' of Einstein's equations (using power series expansions), introducing, instead, fresh ideas from causal structures and behavior of geodesics, assuming that curvature is caused by matter satisfying suitable energy conditions. This fresh perspective provided novel tools to show that singularities would be ubiquitous in general relativity if conditions that are natural in classical physics are met. Considerations of the last three sections suggest that, from a broader perspective that combines general relativity with quantum physics, singularities are harmless if one probes them with physically appropriate tools, or even absent altogether if we take into account the quantum nature not only of matter but also of geometry. But the detailed analysis has been carried out only in symmetry reduced systems. Thus, in a sense, progress has been `helical' and we are back at the question of whether the situation in these examples is generic, albeit at a higher level that includes quantum physics in addition to general relativity. What happens \emph{physically} in generic situations in absence of symmetries? Are the results obtained in symmetry reduced models providing correct pointers? That is: Are generic space-like singularities naturally resolved in quantum gravity? Can one establish `no-singularity theorems' using only appropriate physical conditions that take into account both gravity and the quantum? Do we need a fresh perspective and a novel toolkit to establish this?

Paddy would have been excited by these profound challenges.

\section*{Acknowledgments}
This work was supported by the NSF grants PHY-1806356, the Eberly Chair funds of Penn State, and the Alexander von Humboldt Foundation.

\section*{Data Availability}
This is a mathematical physics paper. No data was collected or used.

\end{document}